\documentstyle[12pt]{article}
\begin{document}

\baselineskip 20pt        
\noindent
\hspace*{13cm}
hep-ph/9808419\\
\noindent
\hspace*{13cm}
KIAS-P98018\\
\noindent
\hspace*{13cm}
YUMS 98-016\\  
\noindent
\hspace*{13cm}
SNUTP 98-094\\ 
\hspace*{13cm}
Brown-HET-1138 \\ 

\vspace{1.8cm}

\begin{center}
{\Large \bf  Neutrino Oscillations and Lepton Flavor Mixing}\\

\vspace*{1cm}

{\bf Kyungsik Kang $^{a,b,}$\footnote{kang@het.brown.edu, kskang@kiasph.kaist.ac.kr}},~~
{\bf Sin Kyu Kang $^{b,}$\footnote{skkang@kiasph.kaist.ac.kr, skkang@supy.kaist.ac.kr}},~~
{\bf C. S. Kim $^{b,c,}$\footnote{kim@cskim.yonsei.ac.kr,~ 
http://phya.yonsei.ac.kr/\~{}cskim/}},
{\bf Sun Myoung Kim $^{d,}$\footnote{skim@hit.halla.ac.kr}}\\

\vspace*{0.5cm}

 $a:$ Department of Physics, Brown University, Providence, RI 02902, U.S.A. \\
 $b:$ School of Physics, Korea Institute for Advanced Study, Seoul 130-012, Korea \\
 $c:$ Physics Department, Yonsei University, Seoul 120-749, Korea \\ 
 $d:$ Halla Institute of Technology, Wonju, Korea \\ 
       
\vspace{1.0cm}
 
(\today)
\vspace{0.5cm}

\end{center}        

\begin{abstract}
\vspace{0.2cm}

\noindent
In view of the recent announcement on non-zero neutrino mass
from Super-Kamiokande experiment, it would be very timely to
investigate all the possible scenarios on masses and mixings of
light neutrinos.
Recently suggested mass matrix texture for the quark CKM mixing,
which can be originated from the family permutation symmetry and
its suitable breakings, is
assumed for the neutrino mass matrix and determined by the four
combinations of solar, atmospheric and LSND neutrino data and 
cosmological hot dark matter  bound as input constraints.
The charged-lepton mass matrix is assumed to be diagonal so that
the neutrino mixing matrix can be identified directly as the
lepton flavor mixing matrix and no CP invariance violation
originates from the leptonic sector.
The results favor hierarchical patterns for the neutrino masses,
which follow from the case when either solar-atmospheric data 
or solar-HDM constraints are used.

\end{abstract}

\vspace{0.3cm}
PACS number(s): 14.60Pq, 12.15Fk

\newpage
Although neutrino mass is predicted to be zero within the minimal Standard Model, 
the solar \cite{solar,superkam} and atmospheric neutrino \cite{superkam,atm} 
observations and the Liquid Scintillator Neutrino Detector (LSND) 
experiment \cite{lsnd} provide likely evidence that neutrinos may have 
nonzero masses and oscillate.
A variety of massive neutrino scenarios \cite{mass,mass2,mass3,mass4,rge} have been 
proposed so as to accommodate the above experimental observations.
Among them, the works \cite{mass2,mass3} that use the
mass matrix ansatz \cite{fritz}  based on symmetry principle 
have attracted much attention.
Mass matrix ansatz, in fact, has been assumed to predict the entire 
Cabibbo-Kobayashi-Maskawa (CKM) matrix in the quark sector, but most of the earlier works
are ruled out \cite{kang1} because they used the simplest texture for
the quark mass matrix and could not predict a heavy
top quark mass as measured at Tevatron \cite{top}.

Recently, a new class of quark mass matrices \cite{kang2,xing} has 
been suggested so that it can be compatible with the measured
top quark mass and the measured values of the CKM matrix elements.
Its specific form is given by
\begin{eqnarray}
M_H=\left( \begin{array}{ccc}
 0 & A & 0 \\
 A & D & B \\
 0 & B & C \end{array} \right )
\end{eqnarray}
In this paper, we would like to examine if such a type of mass matrix 
can be suitable
for the neutrino sector and accommodate the experimental and/or 
cosmological observations of neutrino.
The form of neutrino mass matrix, in general, needs not be the same 
as that of charged lepton mass matrix.

We propose that charged lepton mass matrix  is taken to be diagonal, 
whereas neutrino mass matrix is given by the form Eq.(1). 
Then, the flavor-mixing CKM matrix in the leptonic sector
becomes coincident with the neutrino 
mixing matrix, and CP violation phase can be rotated away in the Yukawa  couplings. 
Very recently, the authors of Ref. \cite{gupta}  considered similar ansatz 
in view of convenience of diagonalizing mass matrices. However, we would like 
to stress that this is another new ansatz in which the
CP invariance can naturally be
imposed to the leptonic sector.

Following the procedure given in Ref.\cite{kang2}, the neutrino mass 
matrix, which respects in general calculability of the flavor mixing matrix, 
can be written in terms of neutrino mass eigenvalues $m_i$:
\begin{eqnarray}
M^{\nu}_r = \left(\begin{array}{ccc}
 0 & \sqrt{\frac{m_1m_2m_3}{m_3-\epsilon}} & 0 \\
\sqrt{\frac{m_1m_2m_3}{m_3-\epsilon}} & m_2 - m_1 + \epsilon &
w(m_2 - m_1 + \epsilon) \\
0 & w (m_2 - m_1 + \epsilon) & m_3 - \epsilon
\end{array} \right),
\end{eqnarray}
where the parameters $\epsilon$ and $w$ are related to each other, 
and the definition
of parameter $w$ is given in Ref. \cite{kang2}.
Confronting the quark mass matrix ansatz as given by the form (2) 
with the measured values of CKM matrix elements, it turned out that  
the experimentally allowed range of $w$ is $0.97 \leq |w| \leq 1.87$ 
in the leading approximation.
In our analysis, we assume the same range of $w$ for the neutrinos.
 
The real matrix $M^{\nu}_r$ given by Eq. (1) can be diagonalized 
by a real orthogonal matrix 
$R^{\nu}$ so that 
$$
R^{\nu} M^{\nu}_r \tilde{R}^{\nu} = diag (m_1, -m_2, m_3).
$$
Then, the explicit form of  
the lepton flavor mixing matrix, i.e. CKM matrix for leptonic sector 
$V_{lep} = R^{\nu}$, 
is given by 
\begin{eqnarray}
V_{lep} = R^{\nu}=\left( \begin{array}{ccc}
f_1 & f_2  & f_3 \\ & & \\
\frac{f_1m_1}{A}  &\frac{f_2(-m_2)}{A }     & \frac{f_3m_3}{A} \\ & & \\
\frac{f_1Bm_1}{(A*(m_1-C))}  &   \frac{f_2Bm_2}{(A*(m_2+C))} &  
\frac{f_3Bm_3}{(A*(m_3-C))}
  \end{array} \right )
\end{eqnarray}
where
\begin{eqnarray}
f_1 &=& [1+(m_1/A)^2+(Bm_1/(A(m_1-C)))^2]^{-1/2}, \nonumber \\
f_2 &=& [1+(m_2/A)^2+(Bm_2/(A(m_2+C)))^2]^{-1/2}, \nonumber \\
{\rm and}~~~~
f_3 &=& [1+(m_3/A)^2+(Bm_3/(A(m_3-C)))^2]^{-1/2}. \nonumber
\end{eqnarray}
Here, the parameters $A,B,C$ and $D$ are related to the mass eigenvalues by
\begin{eqnarray}
C+D &=& m_1-m_2+m_3, \nonumber \\
A^2+B^2 - CD &=& m_1 m_2 + m_2 m_3 -m_3 m_1, \nonumber \\
{\rm and}~~~
A^2C &=& m_1m_2m_3. \nonumber
\end{eqnarray}

Now, we are ready to determine the elements of the lepton mixing (CKM)
matrix $V_{lep}$ from the experimental constraints of
solar, atmospheric and LSND neutrino observations as well as cosmological bound.
Followings are the experimental constraints on neutrino masses and mixings:
\begin{itemize}
\item 
(A) The {\bf solar} neutrino deficit\footnote{Through out this paper, 
we will take the small mixing MSW solution for extracting the neutrino
mass eigenvalues and lepton flavor mixing parameters.}
can be explained through the MSW mechanism \cite{msw} if 
$4\times 10^{-6} \leq \Delta m^2_{solar} \leq 1.2\times 10^{-5}~\mbox{eV}^2$ 
and $3 \times 10^{-3} \leq \sin^2 2\theta_{solar} \leq 1.1\times 10^{-2}$ (small angle case), 
or $8\times 10^{-6} \leq \Delta m^2_{solar} \leq 3\times 10^{-5}~\mbox{eV}^2$ 
and $0.42 \leq \sin^2 2\theta_{solar} \leq 0.74$ (large angle case), 
and through the just-so vacuum oscillations if 
$\Delta m^2_{solar} \simeq 10^{-10}~\mbox{eV}^2$ and $\sin^2 2\theta_{solar} 
\geq 0.7$ \cite{solarr}.
\item 
(B) The {\bf atmospheric} neutrino deficit can be accommodated if 
$0.0005 \leq \Delta m^2_{atm} \leq 0.006 ~\mbox{eV}^2$ 
and $0.82 \leq \sin^2 2 \theta_{atm} \leq 1.0$ \cite{atmr}.
\item 
(C) The {\bf LSND} data indicates 
$0.27 \leq \Delta m^2_{LSND} \leq 10~\mbox{eV}^2$ 
and $0.05 \leq \sin^2 2\theta_{LSND} \leq 1.0$ \cite{lsnd}.
\item 
(D) On the other hand, if light massive neutrinos provide the hot dark matter ({\bf HDM}) 
of the Universe, one has to impose $\sum |m_{\nu_i}| \sim 6$ eV.
\item
We note that if we consider three generations of neutrino, 
those results (A--C) can be interpreted by $\nu_e \rightarrow \nu_{\mu},
\nu_{\mu} \rightarrow \nu_{\tau}$ 
and $\bar{\nu}_{\mu} \rightarrow \bar{\nu}_e$ oscillations for {\bf solar}, 
{\bf atmospheric} neutrinos and the {\bf LSND} experiment, 
respectively \footnote{The recent experiment by CHOOZ \cite{chooz} disfavors 
$\nu_{\mu} \rightarrow \nu_{e}$ oscillation for atmospheric neutrinos.
In addition, more recent result from KARMEN2 seems to have excluded most of
the region suggested by the LSND experiment \cite{karmen2}.}.
As is well known, it is impossible to construct the lepton flavor mixing matrix 
in the three-generation neutrino scenario that accommodates all (A--D) of
the above three neutrino anomalies and cosmological constraint simultaneously.
Thus, one has to sacrifice one or two conditions among the above four results
to make the best possible combinations to be considered.
\end{itemize}

Although there are many possible combinations in three generation light neutrino
scenario, in this paper based on the observed fermion mass hierarchy and the possible
fermion mass degeneracy,
we take only four combinations of neutrino data 
that  would be the most interesting and distinctive:
\begin{itemize}
\item (I) solar-atmospheric-HDM neutrino, 
\item (II) solar-atmospheric neutrino, 
\item (III) solar-HDM neutrino, 
\item (IV) atmospheric-LSND neutrino.
\end{itemize}
We note that those four cases fall into two categories.
One is almost degenerate three-neutrino scenario and the other is 
hierarchical neutrino scenario. Almost degenerate three-neutrino scenario 
can be achieved by choosing the neutrino data set (I), 
while the hierarchical neutrino structures 
are led to by the sets (II), (III) and (IV).
Notice that the sets (II) and (III) have hierarchical structure $m_1 << m_2 << m_3$, 
while the set (IV) has $m_1 << m_2 \sim m_3$.
We also note that the other possible combinations give not much different
physics results from the above four cases (I--IV).

{}From the constraints of neutrino data sets (I) - (IV), 
one can easily extract the appropriate eigenvalues of neutrino masses 
$m_{\nu_i}$: \\
(I) solar-atmospheric-HDM  
\begin{equation}
m_{\nu_1} = 1.9979 ~-~ 1.9996, ~~~ m_{\nu_2} = 1.9979 ~-~ 1.99996, ~~~ 
m_3 = 2.00084~-~ 2.0042 ~~~(\mbox{eV})
\end{equation}
(II) solar-atmospheric 
\begin{equation}
m_{\nu_1} \sim 0 , ~~~ m_{\nu_2} \sim 10^{-2} - 10^{-3},
~~~ m_{\nu_3} \sim 0.0717  ~~~(\mbox{eV})
\end{equation}
(III) solar-HDM              
\begin{equation}
m_{\nu_1} \sim 10^{-5}, ~~~ m_{\nu_2} \sim 10^{-2}, ~~~m_{\nu_3} \sim 3 
 ~~~(\mbox{eV})
\end{equation}
(IV) atmospheric-LSND
\begin{equation}
m_{\nu_1} \sim 0.5\times 10^{-5}, ~~~m_{\nu_2} \sim 0.52,  ~~~m_{\nu_3} \sim
0.524  ~~~(\mbox{eV})
\end{equation}
Note that  we have not introduced any specific model in Eqs. (4--7), 
which are most reasonable and distinctive
based on items (A--D), as explained before.

Following the numerical analysis of Eqs. (2--3) and using the above neutrino
mass eigenvalues (4--7), 
we can obtain the following lepton flavor mixing (CKM) matrices respectively:
\begin{eqnarray}
({\rm I})~~~~ & & V_{lep} = \left(\begin{array}{ccc}
       0.6925 & 0.7071 & 0.1433 \\
       0.6924 & -0.7071 & 0.1434 \\
       0.2027 & -0.00006  & -0.9792 \end{array}\right) \\
({\rm II})~~~~& &  V_{lep} = \left(\begin{array}{ccc}
       0.9950 & 0.0995 & 0.0000 \\
       0.0729 & -0.7293 & 0.6803 \\
       -0.0677 & 0.6769  & 0.7330 \end{array}\right) \\
({\rm III})~~~~& &  V_{lep} = \left(\begin{array}{ccc}
       0.9950 & 0.0995 & 0.0000 \\
       0.0729 & -0.7292 & 0.6804 \\
       -0.0677 & 0.6770  & 0.7328 \end{array}\right) \\
({\rm IV})~~~~& &  V_{lep} = \left(\begin{array}{ccc}
       0.9998 & 0.0198 & 0.0086 \\
       0.0147 & -0.9164 & 0.3999 \\
       0.0158 & -0.3997  & -0.9165 \end{array}\right)
\end{eqnarray}

Now, let us discuss if those obtained lepton flavor mixing (CKM) matrices, Eqs. (8--11),
can be compatible with the constraints of neutrino experiments (items A--D).
The mixing parameters $\sin^2 2\theta$ for the solar, atmospheric and LSND
neutrino oscillations can be related to \cite{bb3}
\begin{eqnarray}
\sin^2 2\theta_{solar} &\approx &4|V_{e1}|^2|V_{e2}|^2
, \nonumber \\
\sin^2 2\theta_{atm} &\approx & 4|V_{\mu 3}|^2|V_{\tau 3}|^2, \nonumber \\
{\rm and}~~~ \sin^2 2\theta_{LSND} &\approx & 4|V_{e 3}|^2|V_{\mu 3}|^2. \nonumber
\end{eqnarray}
\begin{itemize}
\item
This case (I) predicts $\sin ^2 2\theta_{solar}\approx 0.969$
which implies maximal mixing between 
$\nu_{e}$ and $\nu_{\mu}$ and small mixing between $\nu_{\mu}$ and $\nu_{\tau}$.
This mixing pattern is ruled out because the recent CHOOZ and Super-Kamiokande
experiments seem to
disfavor $\nu_{\mu}\rightarrow \nu_{e}$ oscillation for atmospheric neutrino
with a large mixing.
Thus, almost degenerate three-neutrino scenario 
can not be achieved in our scheme. 
\item
The flavor mixing matrix (II) leads to $ \sin^2 2\theta_{solar}\approx 0.039$
and $ \sin^2 2\theta_{atm} \approx 0.995$ which are consistent with the recent
experimental results for solar and atmospheric neutrino anomalies, which can be 
interpreted as neutrino oscillations.
\item
The case (III) shows similar pattern of mixing matrices as in 
(II), but
predicts somewhat different mixing angle 
between $\nu_{\mu}$ and $\nu_{\tau}$ from the case (IV). 
\item
This mixing matrix (IV) leads to $ \sin^2 2\theta_{atm} \approx 0.537$ and
$\sin^2 2\theta_{LSND} \approx 0.0005$ which can not be compatible with 
the results for neutrino oscillations from atmospheric neutrino and LSND 
experiment.
\end{itemize}
Let us check if our neutrino mass matrix ansatz can be free from the constraints
on neutrinoless double beta decay.
As is well known, nonobservation of the neutrinoless double beta decay provides
neutrino mass bound, $<m_{\nu_e}> =|\sum^{3}_{i=1}\eta_i V^2_{\nu_e i}m_i|\leq
0.45 ~\mbox{eV}^2$ \cite{bb}.
If we impose the CHOOZ and Bugey results \cite{bb1}, 
$<m_{\nu_e}> \leq 3\times 10^{-2}$ eV \cite{bb2}.  
As can be easily expected from our mixing matrices (II)--(III),
our numerical results lie below these bounds.

In view of recent announcement non-zero neutrino masses from 
Super-Kamiokande experiment, it would be very timely to investigate
all the possible scenarios on masses and mixings of three light neutrinos.
Starting from phenomenological four texture
zeros lepton mass matrices, which can be originated from the family
permutation symmetry and its suitable breakings, we investigated the physical
consequences by comparing our theoretical predictions with the solar and
atmospheric neutrino observations, the LSND experiment 
as well as the hot dark matter bound.
We found that our scheme favors hierarchical patterns for the neutrino masses, 
which follow from the case when either solar-atmospheric data 
or solar-HDM constraints are used.

\newpage

\noindent
{\Large \bf Acknowledgments}
\\

\noindent
Two of us (KK and CSK) wish to thank the Korea Institute for Advanced Study for warm
hospitality.
The work of KK is supported in part by the US DOE contract DE-FG02-91ER-40688-Task A. 
The work of CSK was supported 
in part by Non-Directed-Research-Fund made in the program year of 1997,
in part by the CTP, Seoul National University, 
in part by the BSRI Program, Ministry of Education, Project No. BSRI-98-2425,
and in part by the KOSEF-DFG large collaboration project, Project No. 96-0702-01-01-2.

\newpage

\end{document}